\documentclass[12pt,preprint]{aastex}

\newcommand{\hst}{{\sl HST}}
\newcommand{\vis}{$m_{\rm F606W}$}
\newcommand{\nir}{$m_{\rm F814W}$}
\newcommand{\Mvis}{M_{\rm F606W}}

\slugcomment{accepted AJ, 25 April 2005}

\shorttitle{NGC 6791}
\shortauthors{King et al.}

\received{2005 March 15}
\begin{document}

\def\subr #1{_{{\rm #1}}}


\title{Color--Magnitude  Diagrams and Luminosity Functions  down to the
Hydrogen Burning  Limit. III.   A Preliminary {\sl  HST} Study  of NGC
6791\footnote{Based  on  observations with  the  NASA/ESA {\it  Hubble
Space Telescope},  obtained at the Space  Telescope Science Institute,
which is operated by AURA, Inc., under NASA contract NAS 5-26555.}}

\author{Ivan R.\ King\altaffilmark{2}, Luigi R.\ Bedin\altaffilmark{3,4},
            Giampaolo Piotto\altaffilmark{3}, Santi
            Cassisi\altaffilmark{5}, \and Jay Anderson\altaffilmark{6}}

\altaffiltext{2}{Department of Astronomy, University of Washington,
Box 351580, Seattle, WA 98195-1580; king@astro.washington.edu}

\altaffiltext{3}{Dipartimento  di Astronomia, Universit\`a  di Padova,
Vicolo dell'Osservatorio 2, I-35122 Padova, Italy;
piotto@pd.astro.it}

\altaffiltext{4}{European Southern Observatory,
Karl-Schwarzschild-Str.\ 2, D-85748 Garching, Germany; lbedin@eso.org}

\altaffiltext{5}{INAF-Osservatorio Astronomico di Collurania,
via M. Maggini, 64100 Teramo, Italy;
cassisi@te.astro.it}

\altaffiltext{6}{Department of Physics and Astronomy, Mail Stop 108,
Rice University, 6100 Main Street, Houston, TX 77005;
jay@eeyore.rice.edu}

\begin{abstract}
Using \hst\ ACS/WFC images,  we derive the color--magnitude diagram of
the old, metal-rich open cluster  NGC 6791 to nearly 29th magnitude in
$V$,  which  is  the   neighborhood  of  the  hydrogen-burning  limit.
Comparison  with isochrones  leads  to a  discussion  of the  distance
modulus,  the reddening,  and the  age of  the cluster.   By  making a
statistical  correction  for  field  stars  we  derive  a  preliminary
luminosity function,  and a very  tentative mass function.   The white
dwarf sequence is clearly shown,  and has been discussed in a separate
paper.
\end{abstract}


%
\section{Introduction}
%

NGC 6791  is unique among  open clusters.  Not  only it is one  of the
richest  of them  (on the  Sky  Surveys it  is easily  mistaken for  a
globular);  it also  has both  an  unusual metallicity  [Fe/H] =  +0.3
(Stetson, Bruntt, \& Grundahl 2003, S03; Worthey \& Jowett, 2003), and
an anomalously old age of  8--12 Gyr (S03; Chaboyer, Green, \& Liebert
1999).   Its  apparent  distance  modulus $(m-M)_V$  =  13--13.5  (see
discussion  in  \S~4)   allows  us  to  measure  stars   down  to  the
neighborhood  of  the   hydrogen-burning  limit.   We  have  therefore
undertaken a  {\it Hubble Space  Telescope} study of this  cluster (GO
9815) in which images made with  the Advanced Camera for Surveys (ACS)
take us down  to $V \simeq29$.  This study is part  of a more extended
investigation aimed  at studying stars  in a low-mass regime  that has
never been accessible before, and will serve as an important guide for
theories  of the  structure of  low-mass stars.   Our color--magnitude
diagrams (CMDs) will check  the luminosity--radius relation, while the
faint end  of the luminosity  function (LF) can  be used to  check the
mass--luminosity relation.  Our target objects (NGC 6397, M4, $\omega$
Centauri, 47~Tuc,  and NGC 6791)  will allow us  to cover most  of the
globular  cluster  (GC)  metallicity  regime,  with  an  extension  to
supra-solar metallicities thanks to NGC 6791.

Our earlier results showed that while models are able to reproduce the
observed main sequence (MS)  almost down to the hydrogen-burning limit
for metal  poor clusters (see  our study of  NGC 6397 in King  et al.\
1998, Paper I of this series),  theory fails to reproduce the CMD of a
moderate-metallicity GC like M4 (Bedin  et al.\ 2001, Paper II).  This
inability  of  stellar   models  to  reproduce  the  luminosity--color
relation, and  the implausible bends at  the low-mass end  of the mass
functions  (MFs)  that  we derive  from  our  LFs  cast doubt  on  the
reliability  of  the   adopted  theoretical  framework.   However,  as
discussed by Baraffe  et al.\ (1998) and Cassisi  et al.\ (2000), this
occurrence must  be related to current uncertainties  in the relations
between  color   and  effective  temperature  that   are  adopted  for
transferring      stellar     models     from      the     theoretical
$(\log(L/L_\odot),\log{T\subr{e}})$    diagram    to    the    various
observational planes. These uncertainties, which are due mainly to the
poorly  known  contribution  of  different  chemical  species  to  the
opacities of  the stellar atmospheres,  become more serious  at higher
metallicities.

Because NGC  6791 lies in  a rich Galactic field  (at $\ell=70^\circ$,
$b=+11^\circ$), our program includes a second epoch of observation, to
be carried out  in mid-2005, to derive proper motions  that we can use
to separate out  field stars.  The first-epoch images  have yielded so
much interesting  information, however, that  we feel that  they merit
the  report that  is given  here.   In fact,  we are  already able  to
present the deepest  CMD of NGC 6791, with a MS  which extends down to
the hydrogen-burning limit,  and, for the first time,  the white dwarf
(WD) cooling sequence, possibly down  to the faintest WDs in NGC 6791.
This paper  will concentrate on the  presentation of the  data set, on
the data reduction,  and on the main-sequence stars.   The white dwarf
cooling sequence will  be discussed in a separate  paper Bedin et al.\
(2005b).

%
\section{Observational Data and Reductions}
%

Our observations, made with the  Wide Field Channel (WFC) of ACS, were
of a  field centered 1.5 arcmin  NE of the cluster  center.  They were
taken on 16 and 17 July, 2003, through the F606W and F814W filters; in
each filter  the exposure times  were 3 $\times$  30 sec +  2 $\times$
1142  sec +  4  $\times$ 1185  sec.   The images  in  each color  were
dithered by fractional pixels, for better photometry and astrometry.

We made our first measurements on  the {\sf \_FLT} images, which are a
pipeline output that includes corrections for bias and flat field.  We
corrected  pixel values  for geometrical  distortion according  to the
recipe given by Anderson (2002).

Then, in  order to reach  as faint as  possible, we created  a stacked
image in  each color,  as follows.  We  first measured  magnitudes and
positions in each image using the ePSF method described by Anderson \&
King  (2000),  and  we   corrected  these  positions  for  geometrical
distortion using the Anderson  (2002) prescription.  We then chose one
of  the  images  as a  reference  frame,  and  found the  best  linear
transformation between each of the other images and this frame.

Since ACS/WFC  images are undersampled, we  made a stack  in which the
original  50 mas  pixels  are subsampled  by  a factor  of  2 in  each
direction.  This  we accomplished as  follows.  For each pixel  of the
super-sampled  image,   we  calculated   the  position  to   which  it
corresponds in  each of the  distortion-corrected images, and  chose a
pixel  value  derived  from  bi-linear interpolation  in  that  image.
(There  is no  need to  allow for  the fact  that the  new  pixels are
smaller, since the  magnitude zero point will be  calibrated only at a
later  stage.)   We calculated  the  mean  of  these 6  values,  using
residuals  to eliminate  outliers by  sigma-clipping at  the $3\sigma$
level.

We will  refer to the resulting  stacks as super-stacks.   For all but
the  brightest stars  our final  photometry was  performed on  the two
super-stacks, using  DAOPHOT (Stetson 1987) and  ALLSTAR (described in
the DAOPHOT User's Manual), with a PSF based on a Moffat function, and
a quadratic spatial variation.

In attempting to  push photometry to a faint limit,  there is always a
problem  with  false detections  at  the  faint  end.  We  used  three
criteria to eliminate false objects.  First, we did the FIND operation
with a threshold of 3.5 times the sigma of the background, in order to
minimize  the background noise  effects.  Then  we eliminated  as many
false objects as possible by  requiring that the absolute value of the
SHA parameter  of DAOPHOT be less  than 0.2 and that  the positions of
the star  in each of the  two super-stacks agree within  a distance of
0.5  pixel.  Before choosing  the numerical  criteria just  quoted, we
carried  out  photometry  using  a  number  of  other  values  of  the
parameters and  examining the resulting  color--magnitude diagrams; we
adopted the values  that appeared to give the  best compromise between
accepting false stars and losing real ones.

Because stars brighter than \vis\  $\simeq$ 20.5 were saturated in the
long exposures, for them we  used photometry by the method of Anderson
\&  King  (2000)  on   the  short-exposure  images.   The  super-stack
magnitudes were put on the  zero point of the short-exposure images by
comparing  the   magnitudes  of  a  large   number  of  well-measured,
non-saturated stars  that were common  to both sets.  The  zero points
were set by the procedures described in Bedin et al.\ (2005a).

Our final  step was a determination  of completeness as  a function of
magnitude.  This  we did by adding artificial  stars (constructed from
our  PSF,   with  a  realistic   amount  of  noise  added).    In  the
artificial-star tests  38,000 stars,  on a grid  with a spacing  of 42
pixels,  were  added  to  each  of the  super-stacks.   The  grid  was
displaced by a random number  of pixels in each coordinate, the number
ranging  from  zero  to  one  less  than  the  separation.   The  same
displacement was used for each color.

The artificial  stars had  random \vis\ magnitudes  over a range  of 7
magnitudes (23.5 $\leq$ \vis\  $\leq$ 30.5).  Two separate experiments
were carried  out.  In one,  the \nir\ magnitude  was chosen so  as to
place the star  on the ridge line of the  white-dwarf sequence; in the
other, the  stars were on  the ridge line  of the main  sequence.  The
super-stacks containing the artificial stars were processed in exactly
the same  way as the real  super-stacks had been.  The  results of the
artificial star  tests on the MS  are shown in Fig.~1.  The left panel
shows,   in  a   CMD,   the  input   (continuous   line)  and   output
magnitudes. The  right panel of Fig.~1  shows that our  MS star counts
are $>50$\% complete for all magnitudes \vis\ $<28.3$.

%
\section{The Color--Magnitude Diagram}
%

%
\subsection{Our Results}
%

Figure~2 shows our full CMD of NGC 6791 for the $\sim$3200 stars which
remain after  the selection described  in the preceding  section.  The
CMD extends for  15 magnitudes, from a couple  of magnitudes above the
TO down to  \vis\ $\sim$ 29, i.e., $\sim6$  magnitudes fainter than in
S03.

The CMD shows a main sequence with a well-defined turn-off, as well as
several magnitudes  of a white-dwarf  sequence.  As expected  from the
relatively low  Galactic latitude of the  cluster ($b=+11^\circ$), the
entire CMD  is contaminated  by field stars.   As mentioned  above, we
will use our second-epoch images (July 2005) to remove the field stars
by measuring the proper motions of all the stars.  The field stars are
not nearly so numerous, however,  as to obscure the cluster CMD, which
is  clearly  visible  down   to  \vis\  $\sim28$.   According  to  the
artificial star  tests the completeness  of our photometry is  50\% at
\vis\ = 28.34, and reaches 90\% at \vis\ = 27.5.  Indeed, we still see
numerous  faint field  stars, but  the  cluster sequence  seems to  be
dwindling away.

In  Fig.~2 there  is a  clear  indication of  a well-populated  binary
sequence  running  parallel  to  the  MS,  and  of  a  possible  white
dwarf-white dwarf binary sequence, parallel to the white dwarf cooling
sequence.  Both sequences indicate  that the  fraction of  binaries in
this cluster  must be relatively  high.  A more  quantitative analysis
will be done when we will have removed the field stars.

Fig.~2 also contains  two very blue stars that  correspond to subdwarf
candidates  discussed by  Kaluzny \&  Rucinski (1995)  and  S03 (Nos.\
11562         and        12652        in         their        catalog,
http://cadcwww.hia.nrc.ca/stetson/NGC6791/NGC6791.dat). (The other S03
subdwarf candidates  are outside our  field.)  These stars  are likely
cluster members (S03), and their presence in such a metal-rich cluster
is noteworthy.  They might correspond  to very hot HB  stars, implying
some  anomalously high mass  loss along  the red  giant branch,  or an
anomalous composition  (e.g., strong  helium enhancement, as  found in
Piotto et al.\ 2005 for a subsample of stars in Omega Centauri).

%
\subsection{Comparison with the Stetson et al.\ CMD}
%

Unfortunately it is not possible to compare our CMD directly with that
of  S03,  on  account  of  differences  in  the  color  systems.   The
magnitudes of  S03 are in  standard {\sl BVI}, whereas  our magnitudes
are  constrained, by  observational  necessity, to  be  in the  system
defined by the F606W and F814W filters of \hst.

Figure~3 shows the  $V-$ \vis\ and $I-$ \nir\  differences between S03
and  our photometry  as, a  function of  the $V-I$  color.  Comparison
between the  S03 $I$  magnitudes and our  \nir, for stars  measured in
common, shows that  on the average they are  nearly identical, for the
range of nearly  8 magnitudes that is covered, with  a small offset of
$\sim$0.05  magnitude (which  we  believe arises  from the  difference
between  the two  passbands rather  than from  any error  in  our zero
point), and  only a  slight non-linearity; but  there is  no guarantee
that this will continue to  fainter magnitudes, where the stars are so
cool  that their  spectra are  full of  complicated features.   Even a
small difference between the two passbands can have sizable effects.

For the $V$ magnitudes, however,  the situation is very different, and
illustrates a  serious problem, often not properly  taken into account
in photometric  studies based  on \hst\ data.   There is  a pronounced
color equation between the S03  $V$ and our \vis.  Furthermore, it can
be expected that  this color equation will be  strongly non-linear for
the fainter, redder  stars.  That this is so can be  shown by means of
synthetic   photometry,   in   which   theoretical   spectral   energy
distributions  (SEDs) are  integrated  over the  passbands of  various
systems and filters.   Not only does the color  equation between \vis\
and $V$ become grossly  non-linear for redder stars; the non-linearity
depends strongly on metallicity.  (We  plan to devote a separate paper
to the problem of color equations.)  Thus the faint part of our MS has
to stand completely on its own; there is no way of tying it to the S03
MS.

The problem, already  mentioned in the Introduction, is  that even the
best available SEDs are able  to match the lower-MS colors of globular
clusters only  at the  lowest metallicities.  One  of the aims  of our
studies, of  course, is to  provide theoreticians with  actual stellar
colors against which they can test their model atmospheres.

{}From this point  of view it makes no difference that  our CMD is not
in $V$ and $I$; at the metallicity of NGC 6791 the transformation from
the   (\vis,$\>$\vis\  $-$   \nir)  system   to   $(M\subr{bol},  \log
T\subr{e})$  is  no  worse  known  than  is  the  transformation  from
$(V,\>V-I)$, since both are quite unknown.   It is a sad fact that for
faint red  stars of the metallicity of  NGC 6791 no SEDs  at all exist
yet, so comparison  of observation with theory will  have to await the
appearance of the theory.

The real value  of a system such as {\sl UBVRI}  is that observers can
intercompare their observations conveniently  if they all agree to use
that system.  For the  comparison of observation with theory, however,
all systems are equally valid,  because comparison depends only on the
basic  operation  of integrating  the  SEDs  of  the theory  over  the
passbands of the observing system.

%
\subsection{Isochrone Fitting}
%

One of  the driving reasons of  this entire project,  and, in general,
the main reason  for studying the CMD of a  cluster, is for comparison
with the  predictions of  stellar evolution models.   Such comparisons
provide important information on the stellar structure and atmosphere,
and allow  measuring important  cluster parameters, including  the age
and  the mass function.  The most  appropriate way  to perform  such a
comparison   is  to   transform  the   theoretical  tracks   into  the
observational  plane.  These  transformations  become  more  and  more
uncertain for stars  much cooler or much hotter than  the Sun, and, as
noted in the  previous Section, the problem becomes  really severe for
supra-solar  metallicities,  to the  point  that  we  expect that  our
observations will  provide important  new inputs to  the theoreticians
working in this field.

Unfortunately,  there is  an additional  problem that  complicates our
efforts:\  the (often large)  uncertainties in  the reddening  and the
distance  modulus of  the  object observed.   These uncertainties  are
particularly large in the case  of NGC 6791, for which various studies
in the  last 20 years  have found reddening  values $0.09<E(B-V)<0.20$
and true distance modulus $12.6<(m-M)_0<13.6$ (see S03, and references
therein).

Definitive  measurements of the  reddening, distance,  and age  of NGC
6791  are beyond  the scope  of  the present  paper.  However,  taking
advantage of  the shape of the  MS around the turn-off  (TO), we still
can extract  some useful information  from our data  (information that
was also used in the companion paper on the WD cooling sequence [Bedin
et al.\  2005b]).  Figure~4 shows the  same CMDs as  in Fig.~2, zoomed
around  the TO.  In  order to  estimate the  cluster age  and distance
modulus, we  used the theoretical framework  presented by Pietrinferni
et al.\  (2004). We refer  the interested reader  to that paper  for a
detailed discussion of the stellar evolutionary models for both H- and
He-burning  phases.  Here  it is  enough  to note  that these  stellar
models  and  isochrones are  based  on  the  most up-to-date  physical
scenario that is currently available.

The  whole  theoretical  framework   has  been  transferred  from  the
theoretical plane  to the  observational Johnson-Cousins one  by using
the recipes  described by Pietrinferni  et al.\ (2004), while  for the
ACS system  on board \hst\ we  adopted the color-effective-temperature
relations,  bolometric-correction  scale,  and  color  transformations
presented by Bedin et al.\ (2005a)\footnote{All the theoretical models
adopted in present work as well as a more extended set of evolutionary
results    and    isochrones    can    be    found    at    the    URL
http://www.te.astro.it/BASTI/index.php.}.  In order  to fit the CMD of
NGC 6791,  we have used  the evolutionary models corresponding  to the
chemical composition $Z=0.03$, $Y=0.288$ (i.e., [M/H] = 0.26).

In Figure~4  we show  the best fit  we could  get by eye.   This would
imply  an  age of  $\sim$9  Gyr,  an  $E(B-V)=0.15$, and  an  apparent
distance  modulus $(m-M)_V=13.5$,  which  would correspond  to a  true
distance modulus $(m-M)_0=13.0$, i.e.,  a distance of $\sim$4 kpc from
us.  Note that while the distance modulus differs by only 0.2 mag from
the  distance modulus  found by  S03, the  reddening  is significantly
higher (by 0.06 magnitude).  (We note also that Carney, Lee, \& Dodson
[2005] recently found from  an infrared study $E(B-V)=0.14\pm0.04$ and
$(m-M)_0=13.04\pm0.04$.)  These differences  in the estimated distance
modulus and reddening are due  in part to the combination of different
assumptions  adopted  in computing  the  stellar  models  used in  the
present work and in the models  used by S03 (for instance, the initial
He content  and physical inputs such  as the equation of  state; for a
detailed discussion of a variety  of stellar models, see VandenBerg et
al.\  2000,  and  Pietrinferni et  al.\  2004),  and  in part  to  the
different bolometric corrections used  to transfer the various sets of
stellar models  from the theoretical plane to  the observational ones.
However, as  shown in  Figs.~5 and 6,  using our  isochrones, properly
transformed into the standard {\sl  BVI} system, with the data of S03,
we obtain an  age, distance, and reddening consistent  with the values
used in  the fit of Fig.~4.   Only with a higher  metallicity ([M/H] =
+0.4) would we obtain  a smaller reddening ($E(B-V)=0.12$, still above
their 0.09).

We also performed  the same exercise as in  Fig.~22 of S03. Basically,
we  compared the  color of  the  TO in  a $B-V$  vs.\ $V-I$  two-color
diagram with  theoretical TO  colors for four  different metallicities
([M/H] = $-$0.25, 0.05, 0.25, 0.4)  and for ages in the interval 7--15
Gyr,   at    steps   of   1    Gyr.    Adopting   a    reddening   law
$E(V-I)/E(B-V)=1.35$, we obtained a larger age ($\sim$12--13 Gyr), and
a   smaller  reddening   ($E(B-V)=0.09$)  than   those   suggested  by
Figs.~4--6, but very similar to  the values suggested by S03.  In this
case one works only with differential numbers within the CMD; however,
the  results  are   based  on  the  (uncertain)  color   of  only  one
observational  point, namely the  TO. We  prefer the  more traditional
approach of isochrone fitting, as it takes advantage also of the shape
of the MSTO and SGB regions.   Indeed, Figs.~4, 5, and 6 indicate that
the shape of the TO-SGB region is well reproduced by an isochrone at 9
Gyr, and that isochrones for ages  $>$10 Gyr are too flat and have too
short a distance between the TO and  the bend in the SGB to be able to
reproduce  the  observed sequence.  In  conclusion,  according to  our
models, both our  CMD on the \hst\ ACS system and  the S03 CMD suggest
that NGC 6791 is  a very old cluster, with an age  greater than 8 Gyr,
assuming for it a metallicity $0.2<[{\rm M/H}]<0.4$.

There are two important notes to add here. First of all, there is some
inconsistency in  the reddenings obtained  in the fits of  Figs.~5 and
6. This  is in  part related  to  the intrinsic  uncertainties of  the
isochrone  fitting  (because  of  the well-known  degeneracy  in  age,
distance, and reddening), but  problems with the transformation of the
models from the theoretical plane to the observational {\sl BVI} plane
cannot be exluded.

More importantly,  Fig.~4 (which is  in the ACS  observational system)
shows that  the models  are unable to  reproduce the observed  CMD for
$m_{\rm  F606W} >19.5$,  being  too blue  compared  with the  observed
sequences.  The same  procedure does give a good fit  to the {\sl BVI}
magnitudes of  S03---clear evidence that the disagreement  here is due
to uncertainty in  the color-effective-temperature relations.  In this
context we should  note that our isochrones were  transferred from the
theoretical diagram  to the standard Johnson system  by using accurate
tabulations of bolometric  corrections and color-$T\subr{e}$ relations
based on the  updated model atmospheres by Castelli  \& Kurucz (2003).
In  the regime  of low-mass  main-sequence stars,  however,  (i.e., MS
stars  with  $T\subr{e}\leq4750$~K),  the empirical  color-$T\subr{e}$
relationships provided  by Houdashelt,  Bell, \& Sweigart  (2000) were
adopted, since  they provided  a better  fit of the  MS locus  of some
selected open  clusters (see Pietrinferni  et al.\ 2004  for details).
In the  case of  the ACS  observational system, we  rely fully  on the
theoretical  color-$T\subr{e}$ transformations  provided  by Bedin  et
al.\ (2005a),  since empirical  calibration is lacking.  This explains
the apparent inconsistency---given that the theoretical stellar models
are  the  same  in  all  the  figures---of  the  results  obtained  in
Figs.~4--6.

For the benefit of comparisons with theory, we quote the colors of the
main-sequence ridge line.  For steps of  0.5 mag in \vis\ from 18.5 to
28.5, the values of \vis\ $-$  \nir\ are 0.76, 0.84, 0.93, 1.05, 1.18,
1.36,  1.53, 1.72,  1.87, 2.03,  2.16, 2.28,  2.39, 2.47,  2.54, 2.62,
2.72, 2.82, 2.94, 3.14, 3.38.

%
\section{Luminosity Function}
%

One of  the ultimate aims  of this project  is to derive  a luminosity
function (LF)  of NGC  6791, with the  field stars cleanly  removed by
means of a  proper-motion study.  At the present  time, even with only
one epoch,  we can  derive a preliminary  LF, by making  a statistical
allowance for the  field stars.  In a CMD of the  cluster, we mark out
the main-sequence  ridge line (MSRL),  and we subtract its  color from
the color of each star, so as to have a MS that stands vertically.  In
this verticalized  CMD it is  fairly easy to  mark a color  range that
comfortably encompasses the MS stars,  with a spread that increases at
fainter  magnitudes  to  allow  in  a  rough  way  for  the  increased
photometric errors.   For the  field stars we  choose a region  to the
blue of the MS, and somewhat separated  from it.  (We do not go to the
red side,  because there  is a strong  indication of a  cluster binary
sequence on that side.)  The color  range for the field stars is twice
as wide,  because the Poisson  error of the field-corrected  MS counts
will then  be $\sqrt{N\subr{clus}  + N\subr{fld}/4}\,$; the  number of
field stars then makes only a small contribution to the error.

Figure 7 shows the successive stages of this process.  For clarity the
verticalized MS is shown at  two different scalings.  At the far right
is the  LF, after  correction for completeness,  with its  error bars.
The dotted  histogram shows the size  of the correction  that was made
for field stars.

The quantity  of real  interest, however, is  the mass  function (MF).
Since the LF and the MF refer to the same stars, it is clear that
$$ L(\Mvis)\,d\Mvis = -f(m)\,dm, $$
where the notation should be obvious; the minus sign is needed because
$\Mvis$ and $m$ increase in  opposite directions.  Thus we find the MF
simply by multiplying the LF by $-d\Mvis/dm$.  To do this, however, we
need to know the mass-luminosity  relation (MLR) at the metallicity of
NGC   6791---which  is   lacking,  because   no  appropriate   set  of
color-$T\subr{e}$   transformations  and  bolometric   corrections  is
available for very-low-mass stars  with metallicity higher than solar.
To alleviate this problem we used the following work-around.  From our
own set  of stellar  models we evaluated  the difference in  the F606W
magnitude  between   stellar  models  with  the  same   mass  and  age
($\approx10$~Gyr), but different metallicities:\ the solar one and the
metallicity adopted for  NGC 6791, $Z=0.03$.  We did  this in the mass
range 0.5--0.8$m_\odot$;  the mean value  was $\Delta\Mvis\approx0.24$
mag.   We  then  applied  this  constant difference  to  a  solar  MLR
extending to very-low-mass  stars, and used the resulting  MLR for the
conversion.   It   has  to  be   emphasized  that  this  is   a  crude
approximation    to     the    true    MLR     for    high-metallicity
stars. Nevertheless,  we think that for  lack of a  more reliable MLR,
this approach should give us a not-too-gross estimate of the MF of NGC
6791.  The results are  shown in Figure 8, in  which successive panels
show the LF, the MF in  linear units, and the MF in logarithmic units.
The MF  is fairly flat, with  irregularities that are  probably due to
non-optimal  field subtraction.  We  will not  comment on  it further,
because of  the crude  way in  which it was  derived, and  because the
second-epoch observations will lead to a more reliable MF.

One  of the  major purposes  of  this project  was to  study the  main
sequence in the neighborhood of the hydrogen-burning limit.  The large
error bars  clearly make  that impossible at  the present  stage.  The
faintest bin, for example, has  only $9\pm4.4$ stars in it.  But after
we get our second-epoch observations, proper motions will identify the
stars that are  actual cluster members, and our study  will be able to
proceed.  Our intention is to do as we did in the case of M4 (Bedin et
al.\ 2001),  using the  much fainter limiting  magnitude of  the stars
found in \nir\ but not in  \vis\ to show that stars having the cluster
motion drop nearly to zero well above the limit.

\section{Summary and Conclusions}
%

First-epoch   \hst\  observations   of  NGC   6791  have   led   to  a
color-magnitude diagram in which the  main sequence can be followed to
magnitudes  in the  vicinity of  the hydrogen-burning  limit---about 6
magnitudes  fainter  than the  ground-based  CMD  of  Stetson et  al.\
(2003).  In  the region of overlap  the agreement is  good, except for
color  equations  between   our  (F606W,$\,$F814W)  system  and  their
$(V,\,I)$  system.   At   fainter  magnitudes,  uncertainties  in  the
spectral energy distributions make it impossible for us to convert our
magnitudes to $(V,I)$.

Fits to  theoretical isochrones, assuming [M/H] =  +0.26, yield values
$(m-M)_{\rm F606W}=13.5$ and $E(B-V)_{\rm  F606W}=0.15$, at age 9 Gyr.
For lack of  the proper-motion separation of cluster  stars from field
that  we  will  have  at  our  second epoch,  we  make  a  statistical
subtraction of  field stars to get a  preliminary luminosity function.
Lacking also a reliable  mass-luminosity relation at this metallicity,
we  make a heuristic  but reasonable  offset from  the solar-abundance
MLR, and derive a tentative MF, which is rather flat.

\acknowledgments

L.R.B.,  S.C.,  and  G.P.\   acknowledge  financial  support  by  MIUR
(PRIN2002, PRIN2003).  J.A.\ and  I.R.K.\ acknowledge support by STScI
grant GO 9815.

\clearpage

\begin{figure}
\epsscale{1.00}
\plotone{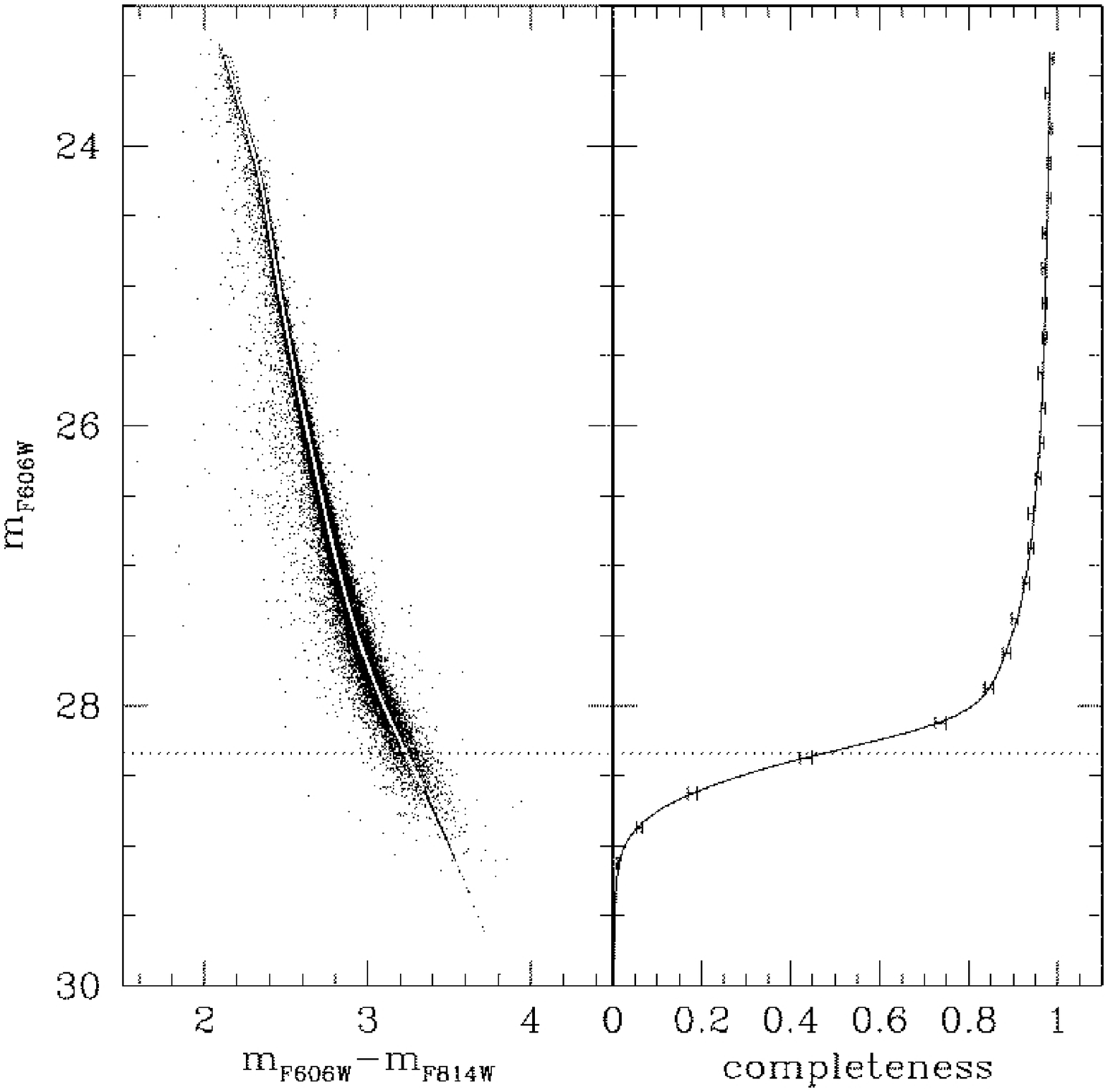}
\caption{Results of the artificial star tests on the cluster MS. The
left panel show the input (continuous line) and output CMD. The
completeness is shown in the right panel.  Our counts have 
completeness 50\% at \vis\ = 28.35.}
\end{figure}

\begin{figure}
\epsscale{1.00}
\plotone{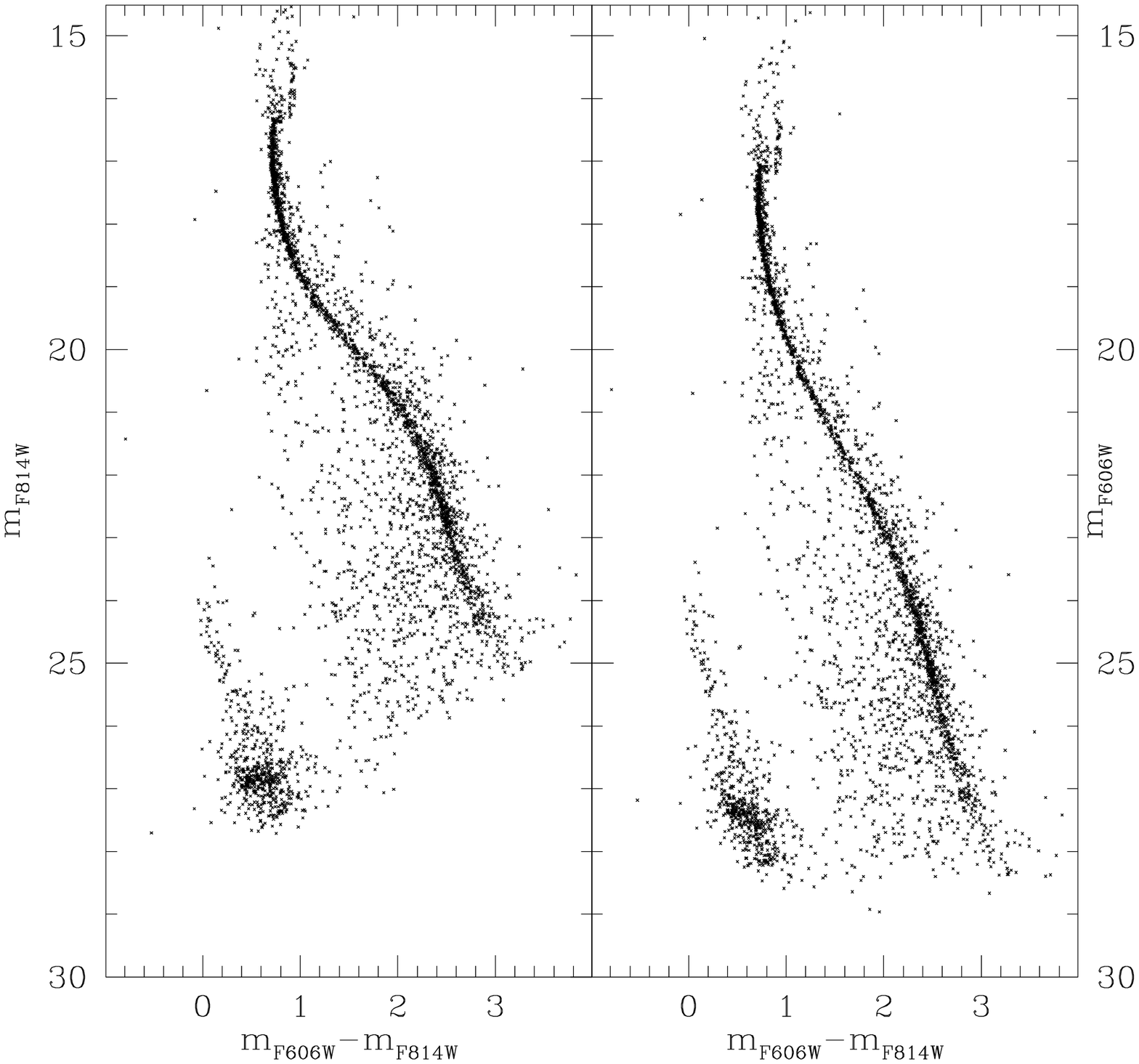}
\caption{The \nir\ vs. \ \vis\ $-$ \nir\ and the \vis\ vs.\ \vis\ $-$
  \nir\ CMDs.} 
\end{figure}

\begin{figure}
\epsscale{1.00}
\plotone{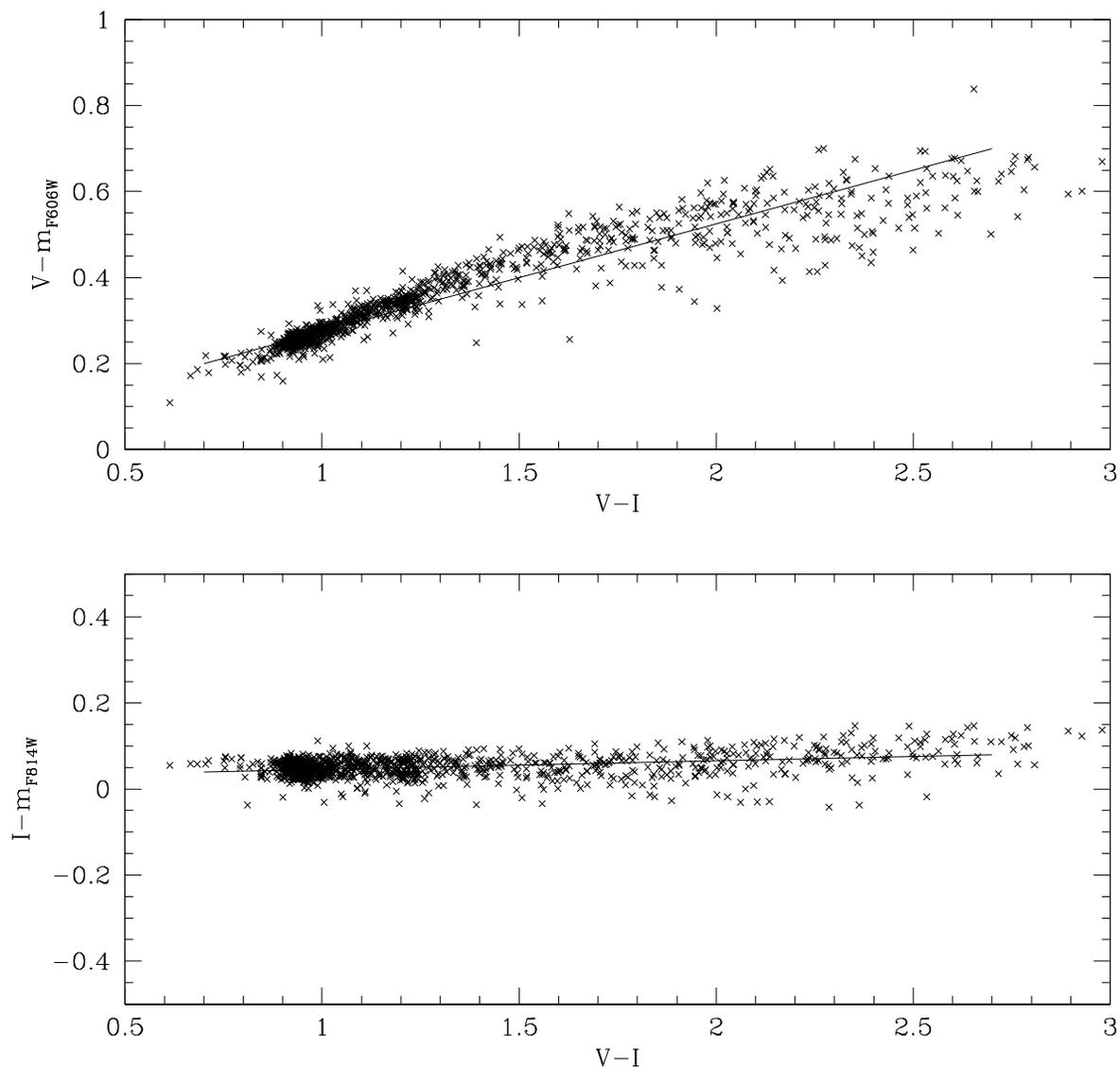}
\caption{The $V-$\vis\ (upper panel) and $I-$\nir\ (lower panel)
differences between S03 and our magnitudes, plotted against the $V-I$
color. The straight lines are drawn arbitrarily, just to make
non-lnearities more evident.  Note the slight non-linearity in the
$I-$\nir\ differences, and the much stronger non-linearity in the
$V-$\vis\ differences.  As explained in the text, the non-linearity will
increase sharply at redder colors.}
\end{figure}

\begin{figure}
\epsscale{1.00}
\plotone{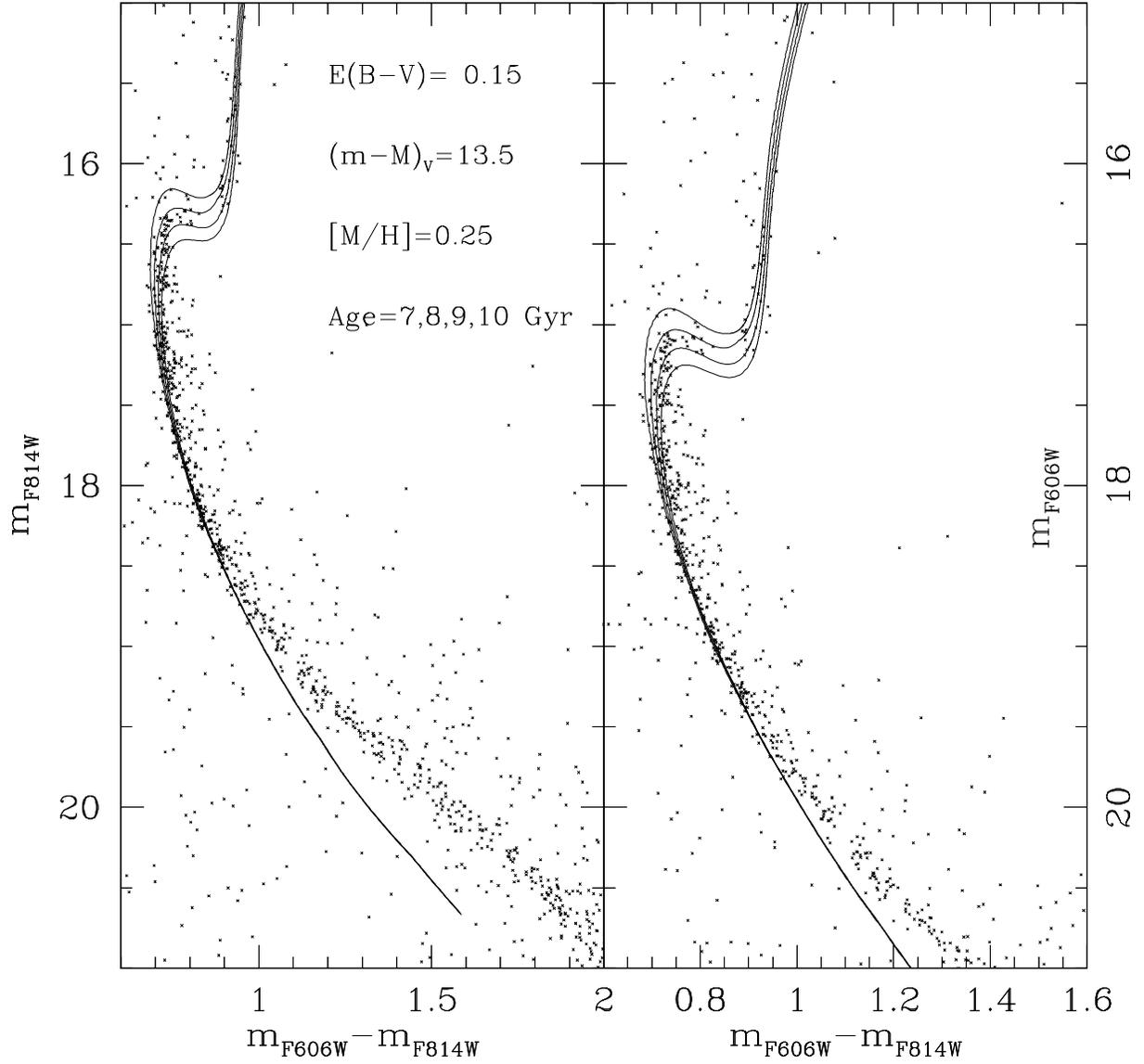}
\caption{The best fit of the Pietrinferni et al.\ (2004) isochrones to the
observed CMD in the ACS observational plane.  It implies an age of 9
Gyr, $E(B-V)=0.15$, and an apparent distance modulus $(m-M)_{F606W}=13.5$.}
\end{figure}

\begin{figure}
\epsscale{1.00}
\plotone{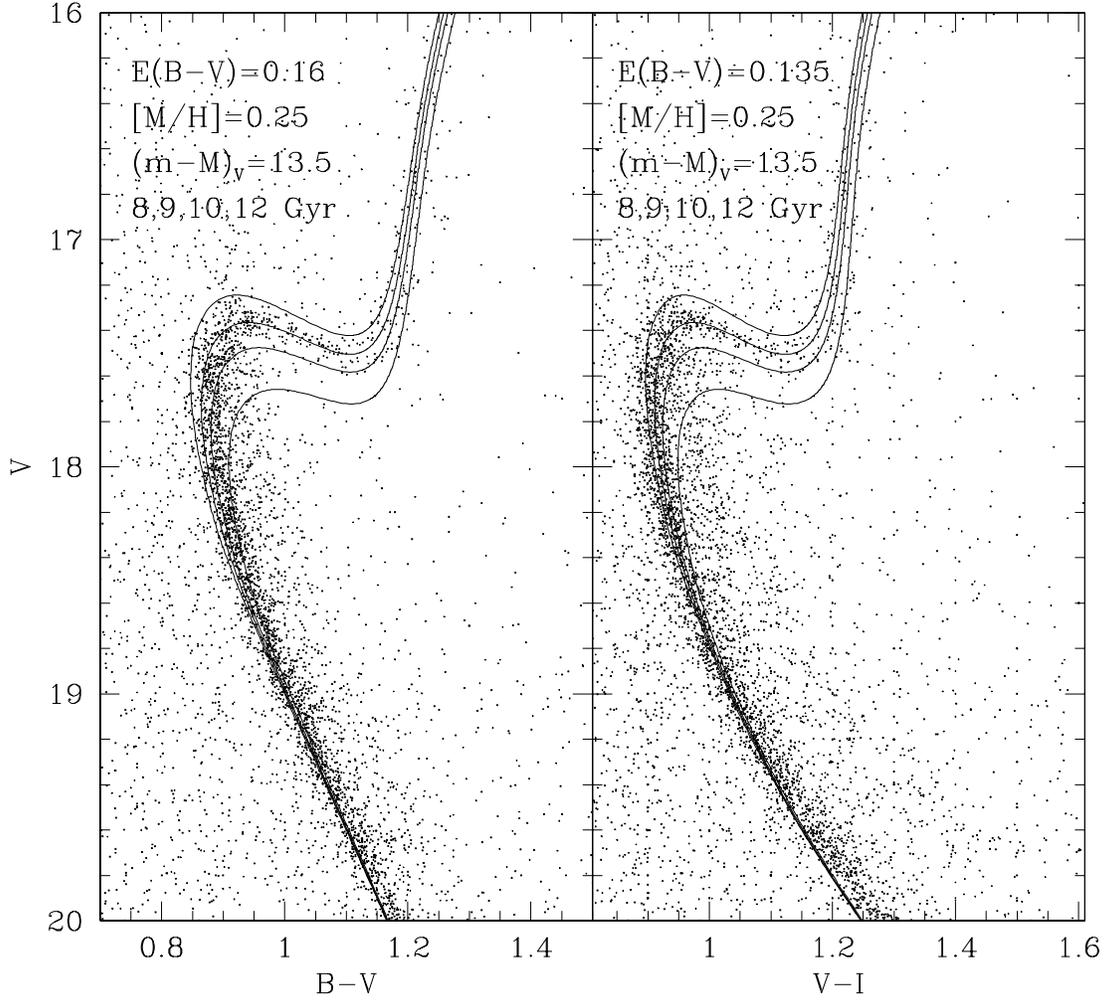}
\caption{Same as in Fig.~4, but here the isochrones are fitted to the
S03 data in the standard {\sl BVI} observational plane.}
\end{figure}

\begin{figure}
\epsscale{1.00}
\plotone{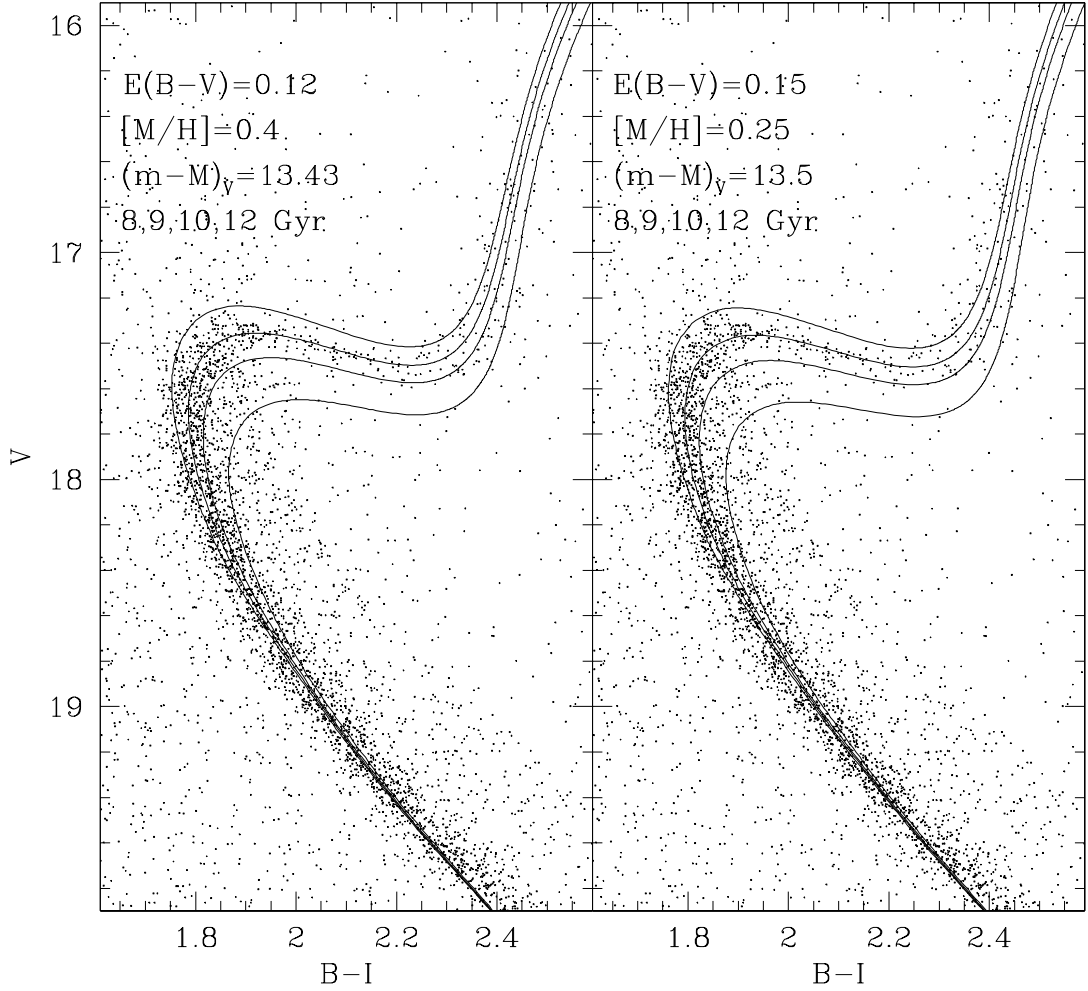}
\caption{Same as Fig.~5, for the $V$ vs.\ $B-I$ CMD (right panel). In the
left  panel the same CMD is fitted with 8, 9, 10, and 12 Gyr isochrones
for [M/H] = 0.4.}
\end{figure}

\begin{figure}
\epsscale{1.00}
\plotone{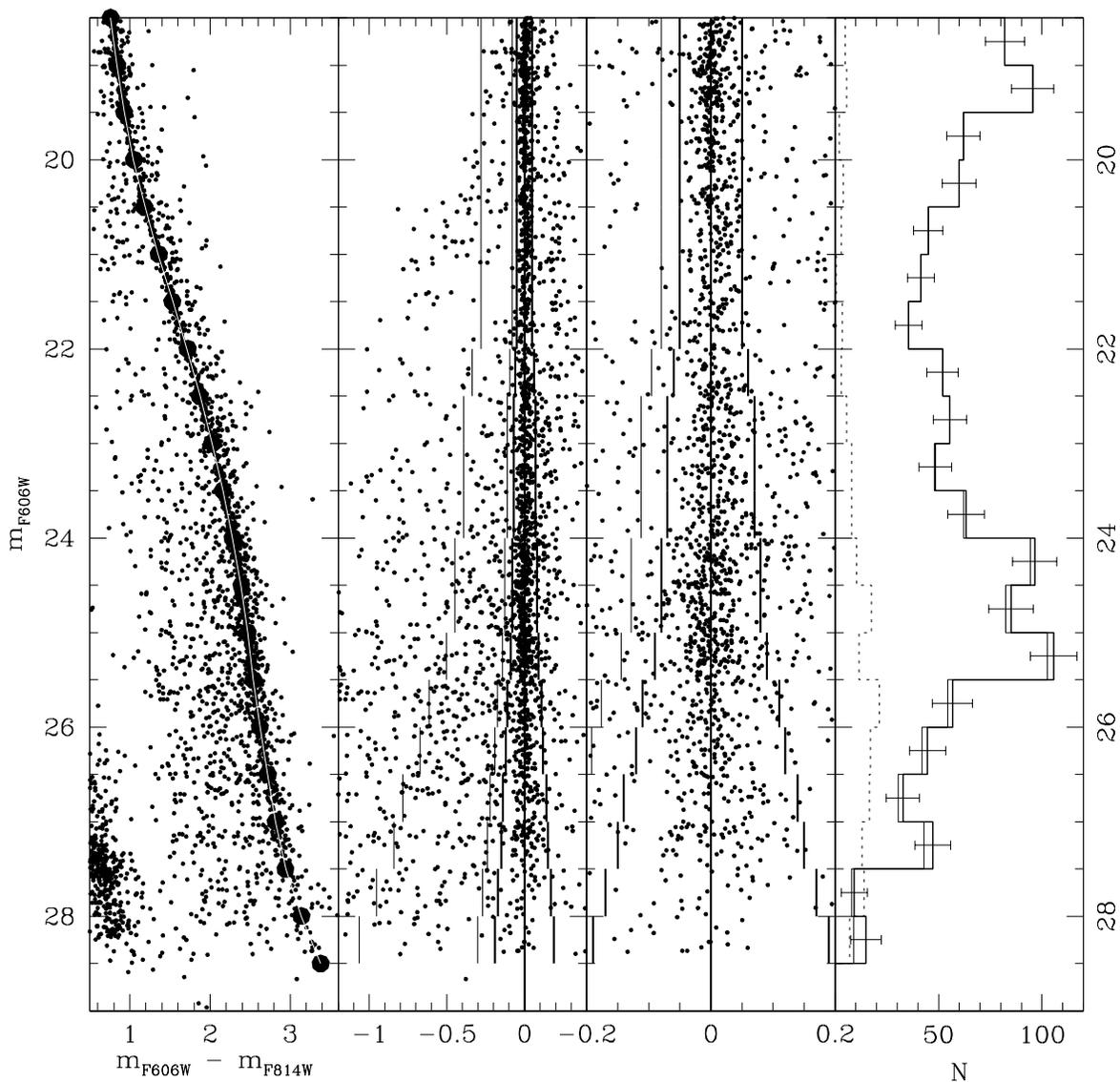}
\caption{The leftmost panel shows the main sequence, with its ridge line
marked by dots; the line through them is a quadratic smoothing.  The two
middle panels show the verticalized MS at two scalings; cluster stars
were counted between the blue lines and field stars between the red
lines.  In the rightmost panel is the luminosity function, before and
after completeness correction; the dotted line shows the correction that
was made for field stars.}
\end{figure}

\begin{figure}
\epsscale{1.00}
\plotone{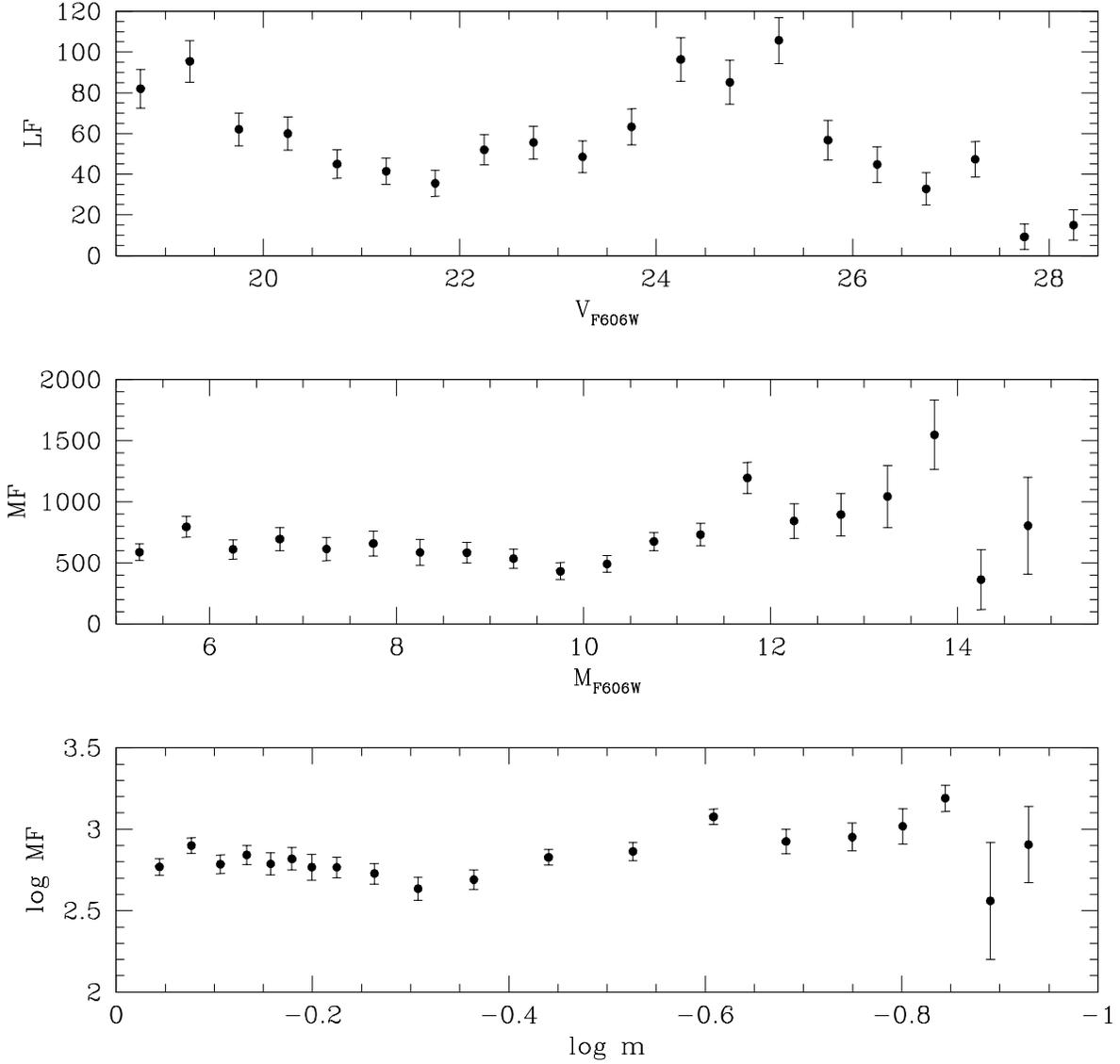}
\caption{The top panel repeats the LF shown in the previous figure.  The
next two panels show the MF, in linear and in logarithmic form.}
\end{figure}

\end{document}